\documentclass[aip,reprint]{revtex4-1}
\usepackage{graphicx}% Include figure files
\begin{document}

% Use the \preprint command to place your local institutional report number
% on the title page in preprint mode.
% Multiple \preprint commands are allowed.
%\preprint{}

\title{ Optically excited multi-band conduction in  LaAlO$_{3} $/SrTiO$_{3}$ heterostructures }

% repeat the \author .. \affiliation  etc. as needed
% \email, \thanks, \homepage, \altaffiliation all apply to the current author.
% Explanatory text should go in the []'s,
% actual e-mail address or url should go in the {}'s for \email and \homepage.
% Please use the appropriate macro for the type of information

% \affiliation command applies to all authors since the last \affiliation command.
% The \affiliation command should follow the other information.

\author{V. K. Guduru}
\affiliation{High Field Magnet Laboratory and Institute for Molecules and Materials, Radboud University Nijmegen, Toernooiveld 7, NL-6525 ED Nijmegen,The Netherlands.}
\author{A. Granados del Aguila}
\affiliation{High Field Magnet Laboratory and Institute for Molecules and Materials, Radboud University Nijmegen, Toernooiveld 7, NL-6525 ED Nijmegen,The Netherlands.}
\author{S. Wenderich}
\affiliation{Faculty of Science and Technology and MESA+ Institute for Nanotechnology, University of Twente, 7500 AE Enschede, The Netherlands.}
\author {M. K. Kruize}
\affiliation{Faculty of Science and Technology and MESA+ Institute for Nanotechnology, University of Twente, 7500 AE Enschede, The Netherlands.}
\author{A. McCollam}
\affiliation{High Field Magnet Laboratory and Institute for Molecules and Materials, Radboud University Nijmegen, Toernooiveld 7, NL-6525 ED Nijmegen,The Netherlands.}
\author{P. C. M. Christianen}
\affiliation{High Field Magnet Laboratory and Institute for Molecules and Materials, Radboud University Nijmegen, Toernooiveld 7, NL-6525 ED Nijmegen,The Netherlands.}
\author{U. Zeitler}
\email{u.zeitler@science.ru.nl}
\affiliation{High Field Magnet Laboratory and Institute for Molecules and Materials, Radboud University Nijmegen, Toernooiveld 7, NL-6525 ED Nijmegen,The Netherlands.}
\author {A. Brinkman}
\affiliation{Faculty of Science and Technology and MESA+ Institute for Nanotechnology, University of Twente, 7500 AE Enschede, The Netherlands.}
\author {G. Rijnders}
\affiliation{Faculty of Science and Technology and MESA+ Institute for Nanotechnology, University of Twente, 7500 AE Enschede, The Netherlands.}
\author {H. Hilgenkamp}
\affiliation{Faculty of Science and Technology and MESA+ Institute for Nanotechnology, University of Twente, 7500 AE Enschede, The Netherlands.}
\author{J. C. Maan}
\affiliation{High Field Magnet Laboratory and Institute for Molecules and Materials, Radboud University Nijmegen, Toernooiveld 7, NL-6525 ED Nijmegen,The Netherlands.}

\date{\today}

\begin{abstract}
The low-temperature resistance of a conducting LaAlO$_{3} $/SrTiO$_{3}$ interface with a 10 nm thick LaAlO$_{3} $ film decreases by more than 50\% after illumination with light of energy higher than the SrTiO$_{3}$ band-gap. We explain our observations by optical excitation of an additional high mobility electron channel, which is spatially separated from the photo-excited holes. After illumination, we measure a strongly non-linear Hall resistance which is governed by the concentration and mobility of the photo-excited carriers. This can be explained within a two-carrier model where illumination creates a high-mobility electron channel in addition to a low-mobility electron channel which exists before illumination.

\end{abstract}

\maketitle

%\section{Introduction}
A conducting interface\cite{Nat04Hwang} between band-insulating perovskites LaAlO$_{3}$ (LAO) and SrTiO$_{3}$ (STO) displays a wide variety of physical phenomena, such as superconductivity,\cite{Scie07Rey} magnetism,\cite{NM07Brink,PRB09Ben,NC11Aria,PRL11Diki,NP11Li,NP11Bert} and quantum oscillations with 2D character.\cite{PRL10Ben,PRL10Cavi,Arix12Mc} Several mechanisms are suggested to describe the origin of conductivity at the LAO/STO interface.\cite{NM06Nak,PRL08Yosh,PRL09Sing,PRL07Siem,PRL07Her,PRL07Will} However, the relative contribution of each mechanism strongly depends on the LAO film thickness and on the LAO growth conditions such as substrate temperature, oxygen partial pressure and the post annealing treatment.\cite{AM09Huij} In particular, growing 5-10 layers of LAO on STO yields a metallic interface with relatively high mobility and low electron concentration,\cite{APL09Bell,PRB10Wong} whereas growing 26 LAO layers with the same conditions results in a low mobility, high concentration electron system with interesting magnetic properties.\cite{NM07Brink} Tuning the transport properties at such a complex oxide interface by modulating the carrier density with light can both contribute to the understanding of its physics and open new pathways towards oxide-based optoelectronic device applications. It has been shown previously that interface conductivity in oxide heterostructures can be tuned by light or by an electric field.\cite{AM10Ras,PRB12Ras,Acs12Teb}

In this Letter we report our investigation of the interface of a LAO/STO sample with 26 monolayers of LAO, using low-temperature (4.2 K) magnetotransport experiments under selective illumination. Illuminating the sample with UV light of energy greater than the STO band gap results in a sharp and persistent decrease of electrical resistance. Using Hall effect measurements, we show that before illumination there is a single, low mobility electron conduction band, and that the resistance drop on illumination can be explained by the creation of a parallel conducting channel containing optically excited high mobility electrons.

%\section{Sample preparation and experimental set-up}
Our sample was grown by pulsed laser deposition and has a 10 nm thick (26 unit cells) LAO film on a TiO$_{2}$-terminated single crystal STO [001] substrate (treatment described in Ref. \onlinecite{APL98Kost}). The LAO film was deposited at a substrate temperature of 850$^\circ$C and an oxygen pressure of $ 2 \times 10^{-3} $ mbar, using a single-crystal LaAlO$_{3} $ target. The growth of the LAO film was monitored using in situ reflection high-energy electron diffraction. After the growth, samples were cooled to room temperature in the deposition pressure.

The sample was mounted on a ceramic chip carrier and electrical contacts were made with an ultrasonic wire-bonder, using aluminium wires. The magnetoresistance and Hall resistance were measured at 4.2 K in van der Pauw geometry, using a standard low-frequency lock-in technique with an excitation current of 1 $\mu$A. The sample was illuminated with light from a broad band Xe-lamp (LSB521), which was filtered with longpass and bandpass filters, and guided through a 30 cm single grating monochromator (Acton-SP2300) to tune the excitation energy. The obtained quasi monochromatic light ($\Delta\lambda$ $\simeq$ 3 nm) was brought to the sample via an optical fiber.

%\section{Experimental results}

\begin{figure}
\includegraphics[width=0.45\textwidth]{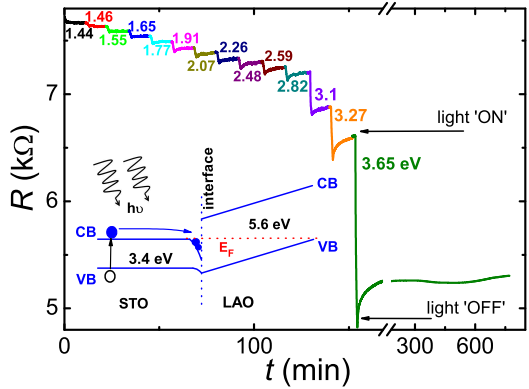}
\caption{Sample resistance as a function of time during the illumination with photons of energy from 1.44 to 3.65 eV at 4.2 K. Each change in the photon energy results in a pronounced step in the sample resistance; the photon energies, in eV, are shown beside each of the steps. The inset shows a schematic band diagram (CB-conduction band, VB-valence band and $E_{F}$-Fermi-level) for a LAO/STO heterostructure under illumination. Note the break on the time-axis showing the persistence of the resistance change}
\end{figure}

In Fig. 1, we show raw data of the sample resistance as a function of time during light illumination in the energy range between 1.44 and 3.65 eV. Each step corresponds to a constant illumination with a specific energy during 1 minute (light~\textquoteleft ON\textquoteright). After illumination, we waited for 10 minutes to allow the resistance to reach a reasonably stable value (light~\textquoteleft OFF\textquoteright), before illuminating with the next photon energy. The influence of photons with different energy is clearly seen as a series of steps in the sample resistance. The largest drop in resistance (more than 50\%), occurs when the photon energy (3.65 eV) exceeds the STO band gap energy (3.4 eV). This dramatically reduced resistance does not recover to the initial value after the illumination is turned off, and only returns to the previous value when the sample is heated to room temperature. We performed a control experiment on a STO[001] substrate under similar illumination conditions, and found that it remains insulating for all photon energies, which proves that the persistent resistance decrease is a feature of the LAO/STO interface.

In Fig. 2, we show as open circles (right axis), the variation in the number of illuminating photons as a function of photon energy, which is a consequence of the energy-dependent throughput of the optical set-up. The number of photons at the sample was calculated from the integration over time of the total power incident on the sample, normalized with the energy of a single photon. The solid circles (left axis), show the resistance change of the sample, normalized by the number of incident photons, at each photon energy. The most drastic change occurs when the photon energy is higher than the STO band gap, shown as the vertical dashed line in the figure.

In order to explain these results, we propose the generation of additional, photo-excited carriers, as depicted in the inset of Fig. 1:\cite{PRB09Son} Illumination with photons of energy greater than the STO band gap creates electron-hole pairs in the STO substrate; the electrons move to the interface potential well, whereas the holes remain trapped in the substrate. Owing to this spatial separation, electron and hole wavefunctions do not overlap and direct optical recombination is suppressed, leading to a persistent resistance change.

\begin{figure}
\includegraphics[width=0.45\textwidth]{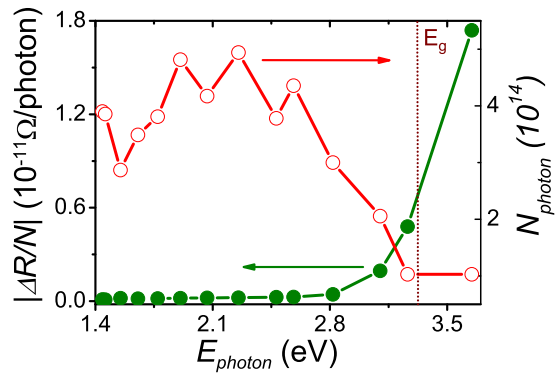}
\caption{ Total number of photons at the sample as a function of photon energy is shown in open circles (right axis). Normalized sample resistance as a function of the energy of illuminating photons is shown in solid circles (left axis). The connecting lines are a guide to the eye.}
\end{figure}

To study the nature of the persistent photo-excited carriers, we have performed magnetotransport experiments after illumination with an increasing total number of photons $N_{tot}$, controlled with neutral density filters, at a constant photon energy of 3.65 eV. In Fig. 3a, we show the Hall resistance as a function of magnetic field at 4.2 K, before illumination and after illuminating with four different values of $N_{tot}$ (open symbols). The corresponding (longitudinal) magnetoresistance are shown in Fig. 3b. Without any illumination, a linear Hall resistance and a small negative magnetoresistance are observed, in agreement with earlier observations on a similar sample.\cite{NM07Brink} After illumination, a distinctly non-linear Hall resistance and a large positive magnetoresistance appear.

We describe the linear Hall resistance using the conventional single-carrier model, and extract the carrier concentration  $(n = B/R_{xy}e)$ and mobility  $(\mu = 1/ \rho_{0} e n)$ from slope of the linear fit to the data - the fit is shown as a solid line in Fig. 3a, where ${B}$ is the applied magnetic field, $ R_{xy}$ is the Hall resistance, $ \rho_{0}$ is the zero field sheet resistance, and ${e}$ is the electronic charge. This yields a carrier density $n_{1}$ = $ 8.9 \times 10^{13} $ cm$^{-2}$ and a mobility $\mu_{1}$ = 3 cm$^{2}$/Vs. This very low carrier-mobility is similar to values previously observed in LAO/STO samples with comparable LAO layer thickness.\cite{NM07Brink,PRB12Her}

\begin{figure}
\includegraphics[width=0.45\textwidth]{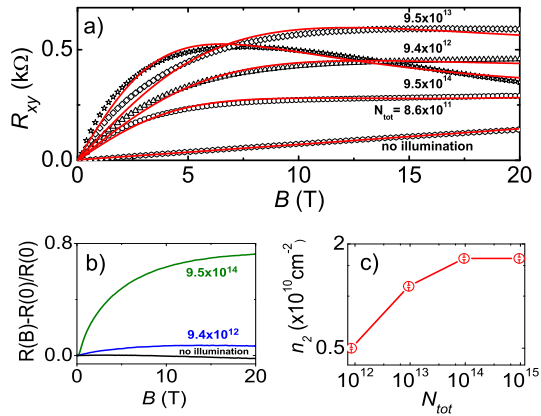}
\caption{(a) Hall resistance data as a function of the applied magnetic field, for illumination with different values of $N_{tot}$ with energy of 3.65 eV at 4.2 K (open symbols). (solid lines) The two-band model fits to the experimental data. (b) (longitudinal) Magnetoresistance data as a function of the applied magnetic field and (c) carrier concentration of the second, high mobility band for illumination with different values of $N_{tot}$.}
\end{figure}

In contrast, the non-linear Hall resistance after illumination cannot be explained within a single-carrier model, but rather suggests a multi-channel system. A similar non-linear Hall resistance was observed previously in LAO/STO,\cite{PRB10Kim,APL10Oht} and explained in terms of two-channel conduction from electronic bands with different mobilities, $\mu_{1}$ and $\mu_{2}$, and carrier densities, $n_{1}$ and $n_{2}$. We use a similar, simple two-electron-band expression for ${R_{xy}}$, given by\cite{BCP76Ash}

\begin{equation}
\ {R_{xy}} = \frac{B}{e}  \frac {({n_{1}\mu_{1}^{2}}+{n_{2}\mu_{2}^{2}}) + (\mu_{1}\mu_{2}{B})^2(n_{1}+n_{2})} {({n_{1}\mu_{1}}+{n_{2}\mu_{2}})^2 + {(\mu_{1}\mu_{2}{B})^2(n_{1}+n_{2})^2}} {,}
\label{eq:one}
\end{equation}
to model our Hall resistance data after illumination.

In this expression, we take $n_{1}$ and $\mu_{1}$ to be the carrier density and mobility of the existing electron band without illumination, and $n_{2}$ and $\mu_{2}$ are the carrier density and mobility of the persistent, photo-excited high mobility band. For low magnetic fields where $\mu_{1}{B}$ $\ll$ 1 and $\mu_{2}{B}$ $\ll$ 1, and for high magnetic fields where $\mu_{1}{B}$ $\gg$ 1 and $\mu_{2}{B}$ $\gg$ 1, expression (1) is linear in magnetic field; R(B) becomes non-linear where $\mu_{2}{B}$ $\simeq$ 1. This simple two-band model is not able to reproduce all the details of the magnetotransport in our sample, but the fact that we observe a non-linear Hall resistance at a few tesla clearly points towards the existence of an optically excited high-mobility channel.

The results of our two-band analysis are shown in Fig. 3a as solid lines. We fitted the non-linear Hall resistance for $N_{tot}$ = $8.6 \times 10^{11}$, using $n_{2}$ and $\mu_{2}$ as fit parameters, and with fixed values of $n_{1}$ = $ 8.9 \times 10^{13} $ cm$^{-2}$ and  $\mu_{1}$ = 3 cm$^{2}$/Vs (as extracted from the linear Hall resistance before illumination): values of $n_{2}$ = $0.5 \times 10^{10}$ cm$^{-2}$ and $\mu_{2}$ = 1200 cm$^{2}$/Vs were obtained. For the higher values of $N_{tot}$ we found that the quality of the fits was insensitive to small variations in $\mu_{2}$ around this value of 1200 cm$^{2}$/Vs. We therefore fixed $\mu_{2}$ = 1200 cm$^{2}$/Vs and used only $n_{2}$ as a fit parameter for the higher values of $N_{tot}$. The values of $n_{2}$ extracted in this way are shown in the Fig. 3c. For the highest value of $N_{tot}$, a good fit to ${R_{xy}}$ required a slightly increased value of $n_{1}$ from $ 8.9 \times 10^{13} $ cm$^{-2}$ to $ 1.19 \times 10^{14} $ cm$^{-2}$, with an unchanged value of $n_{2}$ (as shown in Fig. 3c). This two-band analysis of the Hall resistance strongly suggests that we populate a second high mobility electron channel by illumination above the STO band-gap.

%\section{Summary}
In summary, we have measured magneto-transport in a LAO/STO heterostructure, with a 10 nm thick LAO film, after illumination with selective photon energy. When the photon energy exceeds the STO band gap, the low-temperature resistance decreases by more than 50\% and remains persistent at the lower value. We explain this effect in terms of optical excitation of an additional high mobility electron channel, which is spatially separated from the photo-excited holes, and confirm the presence of a second  conducting electron band through measurement of a strongly non-linear Hall resistance after illumination. A two-carrier description of the Hall resistance data after illumination shows one low mobility (3 cm$^{2}$/Vs) band with a high carrier density ($\simeq$ $10^{14} $ cm$^{-2}$) corresponding to the original conduction band present before illumination, and one persistently photo-excited high mobility (1200 cm$^{2}$/Vs) band with a low carrier density ($\simeq 10^{10}$ cm$^{-2}$).

%\begin{acknowledgments}
This work is part of the InterPhase research program of the Foundation for Fundamental Research on Matter
(FOM, financially supported by the Netherlands Organization for Scientific Research (NWO)).
%\end{acknowledgments}


\begin{thebibliography}{00}
\bibitem{Nat04Hwang} A.~Ohtomo and H.~Y.~Hwang,
 Nature (London) {\bf 427}, 423 (2004).

\bibitem{Scie07Rey}	N.~Reyren, S.~Thiel, A.~D.~Caviglia, L.~F.~Kourkoutis, G.~Hammerl, C.~Richter, C.~W.~Schneider, T.~Kopp, A.~S.~Ruetschi, D.~Jaccard, M.~Gabay, D.~A.~Muller, J.~-M.~Triscone, and J.~Mannhart,
 Science {\bf 317}, 1196 (2007).

\bibitem{NM07Brink}	A.~Brinkman, M.~Van Zalk, J.~Huijben, U.~Zeitler, J.~C.~Maan, W.~G.~Van der Wiel, G.~Rijnders, D.~H.~A.~Blank, and H.~Hilgenkamp,
 Nature~Mater. {\bf 6}, 493 (2007).

\bibitem{PRB09Ben} M.~Ben Shalom, C.~W.~Tai, Y.~Lereah, M.~Sachs, E.~Levy, D.~Rakhmilevitch, A.~Palevski, and Y.~Dagan,
 Phys.~Rev.~B {\bf 80}, 140403 (2009).

\bibitem{NC11Aria} Ariando, X.~Wang, G.~Baskaran, Z.~Q.~Liu, J.~Huijben, J.~B.~Yi, A.~Annadi, A.~R.~Barman, A.~Rusydi, S.~Dhar, Y.~P.~Feng, J.~Ding,  H.~Hilgenkamp, and T.~Venkatesan,
 Nature~Comm. {\bf 2}, 7 (2011).

\bibitem{PRL11Diki} D.~A.~Dikin, M.~Mehta, C.~W.~Bark, C.~M.~Folkman, C.~B.~Eom, and V.~Chandrasekhar,
 Phys.~Rev.~Lett. {\bf 107}, 056802 (2011).

\bibitem{NP11Li} Lu Li, C.~Richter, J.~Mannhart, and R.~C.~Ashoori,
 Nature~Phys. {\bf  7}, 762 (2011).

\bibitem{NP11Bert} J.~A.~Bert, B.~Kalisky1, C.~Bell, M.~Kim, Y.~Hikita, H.~Y.~Hwang, and K.~A.~Moler,
 Nature~Phys. {\bf  7}, 767 (2011).

 \bibitem{PRL10Ben} M.~Ben Shalom, A.~Ron, A.~Palevski, and Y.~Dagan,
  Phys.~Rev.~Lett. {\bf 105}, 206401 (2010).

\bibitem{PRL10Cavi} A.~D.~Caviglia, S.~Gariglio, C.~Cancellieri, B.~Sacepe, A.~Fete, N.~Reyren, M.~Gabay, A.~F.~Morpurgo, and J.~-M.~Triscone,
 Phys.~Rev.~Lett. {\bf 105}, 236802 (2010).

\bibitem{Arix12Mc} A.~McCollam, S.~Wenderich, M.~K.~Kruize, V.~K.~Guduru, H.~J.~A.~Molegraaf, M.~Huijben, G.~Koster, D.~H.~A.~Blank, G.~Rijnders, A.~Brinkman, H.~Hilgenkamp, U.~Zeitler, and J.~C.~Maan
 e-print arXiv:1207.7003[cond-mat.mtrl-sci].

\bibitem{NM06Nak} N.~Nakagawa, H.~Y.~Hwang, and D.~A.~Muller,
 Nature~Mater. {\bf  5}, 204 (2006).

\bibitem{PRL08Yosh} K.~Yoshimatsu, R.~Yasuhara, H.~Kumigashira, and M.~Oshima,
 Phys.~Rev.~Lett. {\bf 101}, 026802 (2008).

\bibitem{PRL09Sing} M.~Sing, G.~Berner, K.~Go\ss{}, A.~M\"uller, A.~Ruff, A.~Wetscherek, S.~Thiel, J.~Mannhart, S.~A.~Pauli, C.~W.~Schneider, P.~R.~Willmott, M.~Gorgoi, F.~Sch\"afers, and R.~Claessen,
 Phys.~Rev.~Lett. {\bf 102}, 176805  (2009).

\bibitem{PRL07Siem} W.~Siemons, G.~Koster, H.~Yamamoto, W.~A.~Harrison, G.~Lucovsky, T.~H.~Geballe, D.~H.~A.~Blank, and M.~R.~Beasley,
 Phys.~Rev.~Lett. {\bf 98}, 196802  (2007).

 \bibitem{PRL07Her} G.~Herranz, M.~Basletić, M.~Bibes, C.~Carretero, E.~Tafra, E.~Jacquet, K.~Bouzehouane, C.~Deranlot, A.~Hamzić, J.-M.~Broto, A.~Barthelemy, and A.~Fert,
  Phys.~Rev.~Lett. {\bf 98}, 216803  (2007).

\bibitem{PRL07Will} P.~R.~Willmott, S.~A.~Pauli, R.~Herger, C.~M.~Schlepütz, D.~Martoccia, B.~D.~Patterson, B.~Delley, R.~Clarke, D.~Kumah, C.~Cionca, and Y.~Yacoby,
Phys.~Rev.~Lett. {\bf 99}, 155502 (2007).

\bibitem{AM09Huij} M.~Huijben, A.~Brinkman, G.~Koster, G.~Rijnders, H.~Hilgenkamp, and D.~H.~A Blank,
 Adv.~Mater. {\bf 21}, 1665 (2009).

\bibitem{APL09Bell} C.~Bell, S.~Harashima, Y.~Hikita, and H.~Y.~Hwang,
 Appl.~Phys.~Lett. {\bf 94}, 222111 (2009).

\bibitem{PRB10Wong} F.~J.~Wong, R.~V.~Chopdekar and Y.~Suzuki,
 Phys.~Rev.~B {\bf 82}, 165413  (2010).

\bibitem{AM10Ras} A.~Rastogi, A.~K.~Kushwaha, T.~Shiyani, A.~Gangawar, and R.~C.~Budhani,
 Adv.~Mater. {\bf 22}, 4448 (2010).

\bibitem{PRB12Ras} A.~Rastogi, J.~J.~Pulikkotil, S.~Auluck, Z.~Hossain, and R.~C.~Budhani,
 Phys.~Rev.~B {\bf 86}, 075127  (2012).

\bibitem{Acs12Teb} A.~Tebano, E.~Fabbri, D.~Pergolesi, G.~Balestrino, and E.~Traversa,
 AcsNANO {\bf 6}, 1278 (2012).

\bibitem{APL98Kost} G.~Koster, B.~L.~Kropman, G.~J.~H.~M.~Rijnders, D.~H.~A.~Blank, and H.~Rogalla,
 Appl.~Phys.~Lett. {\bf 73}, 2920 (1998).

\bibitem{PRB09Son} W.-j.~Son, E.~Cho, B.~Lee, J.~Lee, and S.~Han,
 Phys.~Rev.~B {\bf 79}, 245411  (2009).

 \bibitem{PRB12Her} T.~Hernandez, C.~W.~Bark, D.~A.~Felker, C.~B.~Eom, and M.~S.~Rzchowski1,
  Phys.~Rev.~B {\bf 85}, 161407  (2012).

\bibitem{PRB10Kim} J.~S.~Kim, S.~S.~A.~Seo, M.~F.~Chisholm, R.~K.~Kremer, H.~U.~Habermeier, B.~Keimer, and H.~N.~Lee,
 Phys.~Rev.~B {\bf 82}, 201407  (2010).

\bibitem{APL10Oht} R.~Ohtsuka, M.~Matvejeff, K.~Nishio, R.~Takahashi, and M.~Lippmaa,
 Appl.~Phys.~Lett. {\bf 96}, 192111 (2010).

\bibitem{BCP76Ash} N. W. Ashcroft and N. D. Mermin,
\textit{Solid State Physics} (Harcourt Brace College Publishers, 1976) page number~240.


%\bibitem{PR68Coh} M.~L.~Cohen, and R.~F.~Blunt,
 %Phys.~Rev. {\bf 168}, 929  (1968).

%\bibitem{PRB08Kala} A.~Kalabukhov, R.~Gunnarsson, J.~Börjesson, E.~Olsson, T.~Claeson and D.~Winkler, Phys.~Rev.~B {\bf 75}, 121404 (2007).


\end{thebibliography}
\end{document}